# Terahertz Response of Biological Tissue for Diagnostic and Treatment in Personalized Medicine

N.T. Bagraev[a], L.E. Klyachkin[a], A.M. Malyarenko[a] and K. B. Taranets[b]

[a] Ioffe Institute, St. Petersburg, 194021 Russia
[b] Peter the Great St. Petersburg Polytechnic University, St. Petersburg, 195251 Russia

**Abstract**—A spectrometer based on silicon nanosandwiches (SNSs) is proposed for problems of personalized medicine. SNS structures exhibit properties of terahertz (THz) emitter and receiver of the THz response of biological tissue. Measurements of the *I–V* characteristics of the SNS structure make it possible to analyze the spectral composition of the THz response of biological tissue and determine relative contributions of various proteins and amino acids contained in the structure of DNA oligonucleotides and the corresponding compounds. Evident advantages of the proposed method are related to the fact that the THz response can be directly obtained from living biological tissue and, hence, used for express analysis of the DNA oligonucleotides. Tests of several control groups show that the further analysis of the specific features of the spectral peaks of the SNS *I–V* characteristics is of interest for methods of personalized diagnostics and treatment.

## INTRODUCTION

Terahertz (THz) radiation spans a wide range of frequencies (100 GHz–30 THz) and wavelengths (3 mm–10 μm) belonging to the short-wavelength part of the millimeter wavelength range, entire millimeter range, and far-IR range. THz photons have relatively low energies, so that the radiation is nonionizing. Note that the THz radiation stimulates important biological reactions in spite of the fact that the radiation is attenuated by more than four orders of magnitude at penetration depths from the skin surface of several hundreds of micrometers. All proteins and protein compounds exhibit absorption and emission in the THz range. Unfortunately, THz radiation does not penetrate through the atmosphere, so that only artificial sources can be used. The corresponding devices (e.g., free-electron lasers, travelling wave tubes, and thermal sources of incoherent radiation) are cumbersome and expensive, and low-noise liquid-helium-cooled bolometers are used for detection of the THz radiation. However, recent progress in nanotechnology of semiconductors and high-temperature superconductors has made it possible to fabricate compact solid state devices for THz emission and detection [1, 2]. Thus, experiments can be performed in the recently inaccessible spectral interval, which is promising for practical applications.

THz radiation freely passes through paper, wood, several construction elements, plastics, ceramics, upper skin layers, and clothes. In several European countries, gigahertz waves are used for scanning of passengers and luggage at airports instead of harmful X-ray radiation. THz radiation is also promising for personal identification, since the structure of the DNA oligonucleotides is related to a particular frequency in the THz range of electromagnetic emission (absorption) spectrum. In this work, we propose personal identification based on the analysis of the *I–V* characteristic of a THz emitter under conditions for reflection of the THz radiation from a biological object.

## 1. MATERIALS AND METHODS

The experimental setup consists of a source of the THz radiation, a Keithley 6221 stabilized current source, and an Agilent 34420A nanovoltmeter. Synchronization is provided by the National Instruments Lab View software. The stabilized current is varied in an interval of [–3.5; 3.5] μA with a step of 0.01 μA, 20 measurements are performed at each point, and the results of subsequent averaging are presented at the plots below.

A silicon nanosandwich (SNS) serves as the source of the THz radiation. The SNS represents an ultranarrow *p*-type silicon quantum well (*p*-SiQW) confined by δ barriers that are heavily doped with boron (5 ×10$^{21}$ cm$^{-3}$) on the *n*-Si(100) surface (Fig. 1) [3, 4]. Relatively high mobility of carriers is reached in the structure. The *p*-Si–QW are formed on the *n*-Si(100) substrates in the course of preliminary oxidation and sub-sequent short-term diffusion of boron from gas phase [3, 5]. Boron atoms form trigonal dipole centers($B^+ - B^-$) in the δ barriers owing to the negative-U reaction $2B^0 \rightarrow B^+ + B^-$ [5]. Crystallographically oriented sequences of such centers form edge channels that provide conduction in *p*-Si–QW. The sheet density of holes was determined to be 3 Ч 10$_{13}$ m$_{–2}$ using the field Hall dependences [4]. The *p*-Si–QW edge channels serve as efficient sources of the THz and GHz radiation in the presence of lateral current due to negative-U dipole centers of boron (Fig. 1) [6].

The characteristics of radiation can be controlled using variations in the stabilized source–drain current or microcavities added to the structure and working at different wavelengths. Several experimental methods have been used to prove the THz and GHz emission of the SNSs [3]. Experiments with a Fourier spectrometer have shown that variations in the characteristic of microcavities embedded in the Si–QW edge channels lead to variations in the modulation frequency and depth of the GHz radiation that modulates the THz response of a biological object [3].

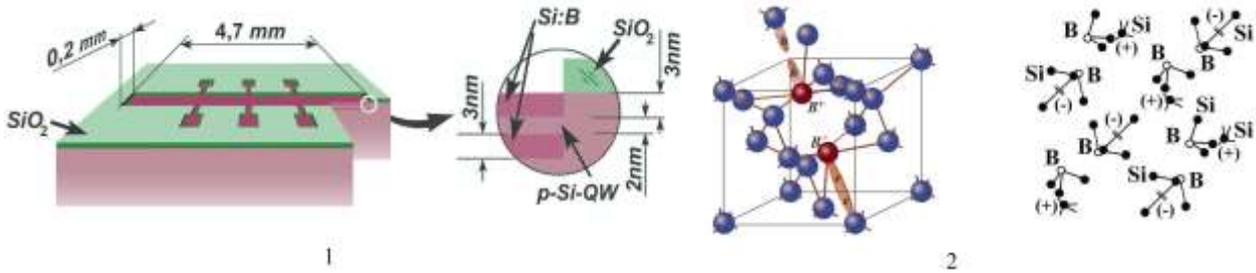

**Fig. 1.** (a) Scheme of the SNS with characteristic sizes and (b) trigonal dipole center ($B^+ - B^-$) with negative correlation energy and chains of the boron dipole centers in the $\delta$ barriers that confine the $p$-Si–QW.

In the experiments, the emitter that also serves as a receiver of the THz–GHz response of a biological object is directed to the area under study being installed at a distance of 1 cm from the surface. Thus, the device works as a balanced photodetector. The generation and detection of the THz–GHz radiation is based on the quantum Faraday effect [6].

## 2. RESULTS AND DISCUSSION

Five volunteer patients took part in the tests. Different organs (left- and right-hand thumbs, breast, and thyroid gland) were studied to reveal reproducibility of frequencies and intensities for a single patient and correlations for the same organs of different patients. We concentrated on a spectral interval of 2–3.5 THz that corresponds to the emission spectra of the DNA oligonucleotides [7] and exhibits specific features of the THz emission of the G–C and A–T pairs (Fig. 2) and the GHz spectrum in the vicinity of a frequency of 160 GHz, which often serves as the modulation frequency of the THz response of biological objects. Note that such a frequency corresponds to the cosmic microwave background radiation.

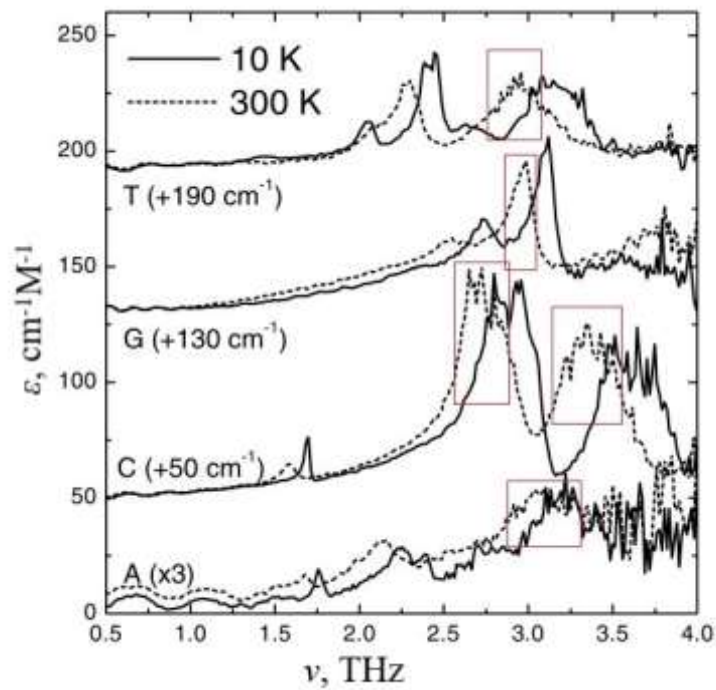

**Fig. 2.** Spectral dependences of the extinction ratio of (A) adenine, (C) cytosine, (G) guanine, and (T) thymine obtained at temperatures of (solid line) 10 and (dashed line) 300 K [7].

Figures 3 and 4 show responses in different spectral intervals of different organs of a single patient. Note developed difference of the signal amplitudes for thumbs, thyroid gland and breast. Significant differences for the thumbs can be due to the difference of the blood vessel structures of the halves of human body. In addition, note characteristic features in the frequency interval corresponding to the response of the DNA oligonucleotides (Fig. 4). A detailed analysis of such an interval will possibly allow determination of the peak positions for dominant C–G and A–T pairs. Thus, the THz responses of different organs of a subject can be used for personal identification.

Figures 5–8 compare the responses of the thyroid gland and thumb. Each plot shows three curves with repeated pattern the nature of which must be interpreted. We assume that such a feature is related to the response of nervous cells. Note also spectral features in the frequency interval of the DNA oligonucleotides (Figs. 6 and 8).

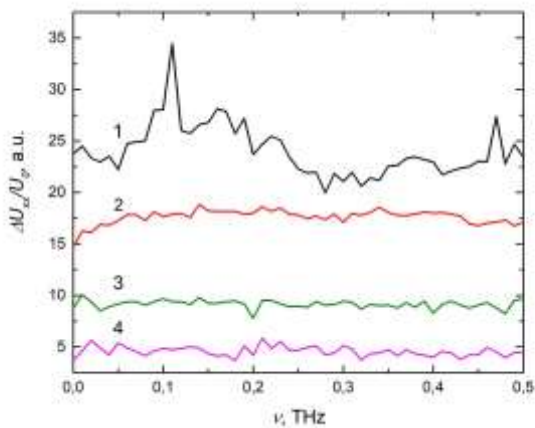

**Fig. 3.** Terahertz responses of (*1*) left-hand thumb, (*2*) right-hand thumb, (*3*) breast (+5 rel. units), and (*4*) thyroid gland of one patient.

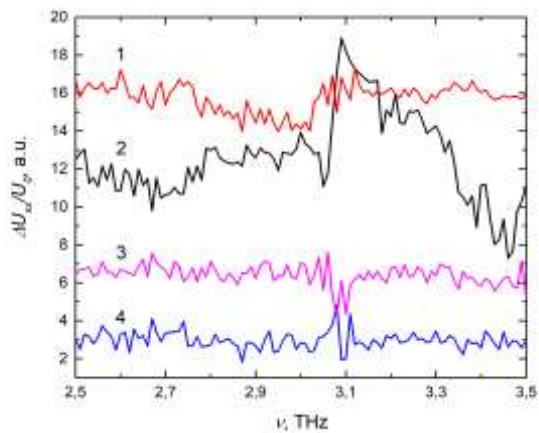

**Fig. 4.** Terahertz responses of (*1*) left-hand thumb, (*2*) right-hand thumb, (*3*) breast (+5 rel. units), and (*4*) thyroid gland of one patient.

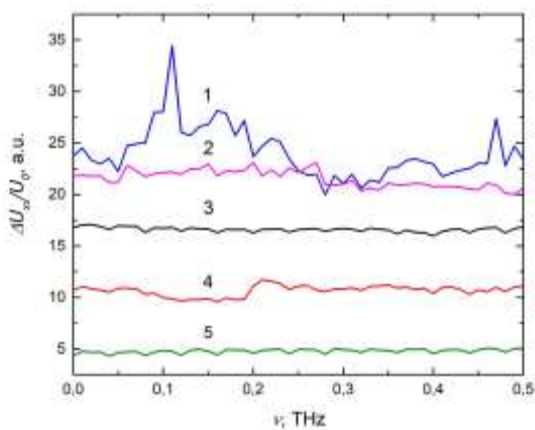

**Fig. 5.** Terahertz responses of left-hand thumbs (*1*, *2*) +15 rel. units, (*3*)+10 rel. units, (*4*) +5 rel. units, and (*5*) of different patients.

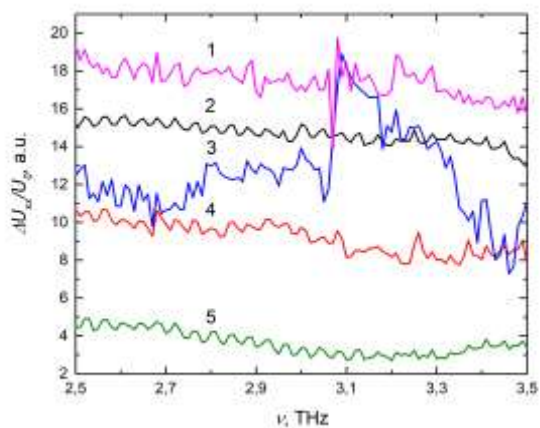

**Fig. 6.** Terahertz responses of left-hand thumbs (*1*) +15 rel. units, (*2*) +10 rel. units, (*3*, *4*) +5 rel. units, and (*5*) of different patients.

## CONCLUSIONS

We have presented an SNS spectrometer that provides emission in the THz spectral range and allows detection of the THz response of biological tissue. The SNS that has been fabricated using planar silicon technology on the *n*-Si(100) represents an ultranarrow *p*-type silicon quantum well confined by δ barriers that are heavily doped with boron ($5 \times 10^{21}$ cm$^{-3}$). In the presence of the source–drain current flow, the edge channels consisting of the boron dipole centers with negative correlation energy serve as efficient sources and receivers of the THz radiation the characteristics of which can be controlled using microcavities embedded in the *p*-Si–QW edge channels.

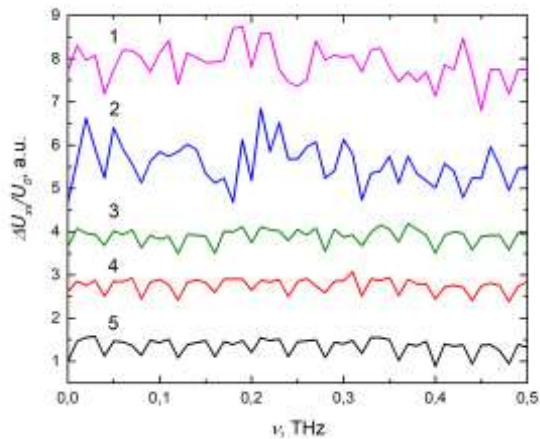 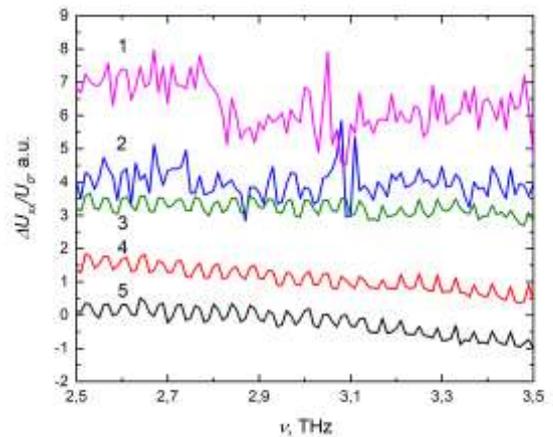

**Fig. 7.** Terahertz responses of thyroid glands (*1*) and (*2*) +10 rel. units, (*3*) +20 rel. units, (*4*) +10 rel. units, and (*5*) of different patients.

**Fig. 8.** Terahertz responses of thyroid glands (*1*) and (*2*) +10 rel. units, (*3*) +20 rel. units, (*4*) +10 rel. units, and (*5*) of different patients.

Tests of control groups have shown that measurements of the *I–V* characteristic of the SNS structure in the presence of the THz response of a biological tissue makes it possible to analyze spectral data and, hence, perform express analysis of relative contributions of proteins and amino acids contained in the DNA oligonucleotide compounds. A further study of the spectral peaks of the SNS *I–V* characteristics is of interest for the methods of personalized diagnostics and treatment.

## FUNDING

This work was supported by the Research Program of the Ioffe Physical Technical Institute, Russian Academy of Sciences.

## COMPLIANCE WITH ETHICAL STANDARDS

All procedures performed in studies involving human participants were in accordance with the ethical standards of the institutional and/or national research committee and with the 1964 Helsinki Declaration and its later amendments or comparable ethical standards.

Informed consent was obtained from all individual participants involved in the study.

## CONFLICT OF INTERESTS

The authors declare that there is no conflicts of interest.